\documentclass{egpubl}
\usepackage{pg2021}

\SpecialIssueSubmission    

\usepackage[T1]{fontenc}
\usepackage{dfadobe}  
\usepackage{color}
\usepackage{amsmath}
\newcommand{\chkA}[1]{#1}
\newcommand{\chkD}[1]{#1}
\newcommand{\chkE}[1]{#1}
\newcommand{\chkF}[1]{#1}
\newcommand{\chkG}[1]{#1}
\newcommand{\chkI}[1]{#1}

\usepackage{tabularx}
\usepackage{wrapfig} 

\providecommand{\norm}[1]{\lVert#1\rVert}

\biberVersion
\BibtexOrBiblatex
\usepackage[backend=biber,bibstyle=EG,citestyle=alphabetic,backref=true]{biblatex} 
\electronicVersion
\PrintedOrElectronic
\ifpdf \usepackage[pdftex]{graphicx} \pdfcompresslevel=9
\else \usepackage[dvips]{graphicx} \fi

\usepackage{egweblnk} 

\title[Relighting Humans in the Wild: Monocular Full-Body Human Relighting with Domain Adaptation]%
{Relighting Humans in the Wild:\\ Monocular Full-Body Human Relighting with Domain Adaptation}

\author[D. Tajima \& Y. Kanamori \& Y. Endo]
{\parbox{\textwidth}{\centering 
        D. Tajima$^{1}$\orcid{0000-0001-7756-0901}
        Y. Kanamori$^{1}$\orcid{0000-0001-5923-423X} 
        Y. Endo$^{1}$\orcid{0000-0001-5923-423X} 
        }
        \\
{\parbox{\textwidth}{\centering $^1$University of Tsukuba, Japan
      }
}
}

\begin{document}

\teaser{
 \includegraphics[width=\linewidth]{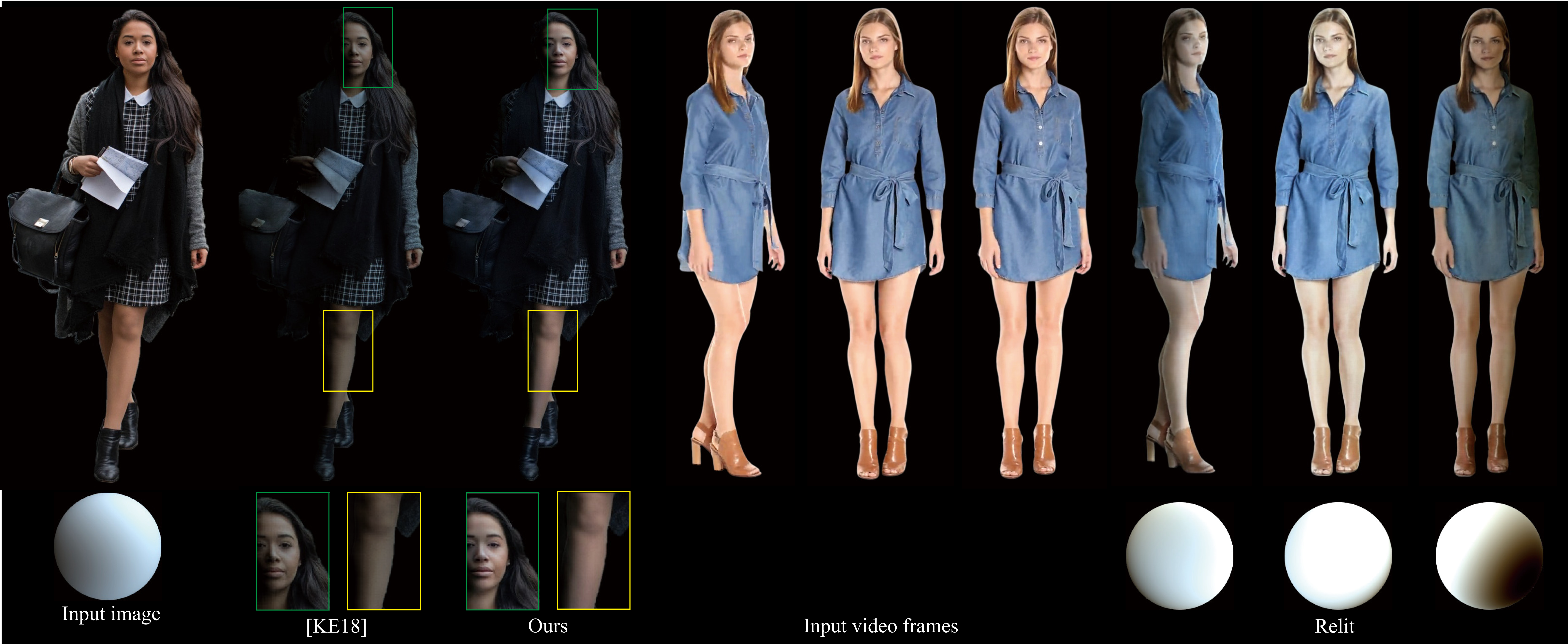}
 \centering
  \caption{\chkE{Given a masked human image, our method can relight the subject with more faithful skin highlights than \chkG{the method by Kanamori and Endo}~\cite{base} (left). We also achieve stable video relighting even under dynamic illuminations (right).}}
\label{fig:teaser}
}

\maketitle
\begin{abstract}
\chkE{The modern supervised approaches for human image relighting rely on training data generated from 3D human models. However, such datasets are often small (e.g., Light Stage data with a small number of individuals) or limited to diffuse materials (e.g., commercial 3D scanned human models). Thus, the human relighting techniques suffer from the poor generalization capability and synthetic-to-real domain gap. In this paper,} \chkD{we propose a two-stage method for single-image human relighting \chkE{with domain adaptation}.
In the first stage, we train a neural network for diffuse-only relighting.
In the second stage, we train another network for enhancing non-diffuse reflection by learning residuals between real photos and images reconstructed by the diffuse-only network.
\chkE{Thanks to the second stage,}
we can achieve higher generalization capability against various \chkE{cloth} textures\chkE{, while reducing the domain gap.} 
Furthermore, \chkE{to handle input videos,} we \chkE{integrate {\em illumination-aware deep video prior} to greatly reduce flickering artifacts even with challenging settings under dynamic illuminations.}}

\ccsdesc[500]{Computing methodologies~Image manipulation}
\ccsdesc[300]{Computing methodologies~Neural networks}

\printccsdesc   
\end{abstract}  
\section{Introduction}

\chkD{Human image relighting allows us to change lighting effects on a photographed person under \chkE{a} new illumination. 
\chkE{We have witnessed rapid, substantial progress in the photorealism of human image relighting, thanks to the deep learning approaches.
The current mainstream of learning-based relighting techniques relies on supervised training with datasets generated from 3D human models.
However, the scalability or quality of the training datasets causes problems. For example, the specialized capturing device called the \textit{Light Stage}~\cite{arfhf} provides high-quality data for relighting, but the Light Stage data are accessible only to self-sufficient research teams. As such, Light Stage datasets are often small (e.g., 70 individuals \cite{TotalRelighting}). On the other hand, we can construct a larger synthetic dataset by collecting 3D scanned human models from commercial websites, but most such 3D models support diffuse materials only. Consequently, human relighting techniques suffer from poor generalization capability against, in particular, various cloth shapes and textures and synthetic-to-real domain gap. }}  

\chkE{To overcome the domain gap while obtaining higher generalization capability, we propose a two-stage supervised method for human image relighting with domain adaptation.}
\chkD{The first stage trains a neural network to infer diffuse-only reflection from the 3D models. 
For this, we extend the network of Kanamori and Endo~\cite{base} with the spatially-frequency (SF) loss~\cite{sfloss} to enhance high-frequency details and a skin mask~\cite{sipre} to emphasize skin regions in estimating albedo and light.
The second stage introduces another network to learn non-diffuse reflection; we apply the diffuse-only network to real photos and reconstruct diffuse-only images. 
The second-stage network then learns the residuals between the real photos and diffuse-only reconstructions.
Because we can collect many 3D models and real photos for training (\chkE{e.g., 521} 3D models and \chkE{8,900} photos in our case), our networks can achieve higher generalization capability against various input data.}

\chkD{We further extend our method to handle human videos.
Na\"{i}vely applying our method frame-by-frame suffers from flickering artifacts.
To enhance the temporal consistency, we integrate the \textit{deep video prior}~\cite{dvp}, \chkE{the state-of-the-art video stabilizer}.
However, we found a problem when we applied dynamic illuminations for video relighting; because the deep video prior stabilizes output frames too strongly, resultant videos do not exhibit time-varying shading effects due to dynamic illuminations. 
By conditioning the deep video prior with the dynamic illuminations, we can synthesize flicker-\chkE{tolerant} relighting videos with dynamic illuminations.
Note that our approach is versatile; it can be applied to any frame-by-frame relighting methods with dynamic illuminations to \chkE{greatly reduce} flickering.}

\chkD{We summarize our contributions as follows:}
\begin{enumerate}
    \item \chkD{A two-stage framework for human image relighting that supports non-diffuse reflection \chkE{with domain adaptation},}
    \item \chkD{An improved diffuse-only inverse rendering with light occlusions, and}
    \item \chkD{An illumination-conditioned deep video prior for flicker-\chkE{tolerant} video relighting with dynamic illuminations.}
\end{enumerate}

For the source code and pre-trained network models, please visit our project page\footnote{\webLink{http://cgg.cs.tsukuba.ac.jp/~tajima/pub/relighting_in_the_wild/}}.




\begin{figure*}[tb]
\centering
\includegraphics[width=16cm]{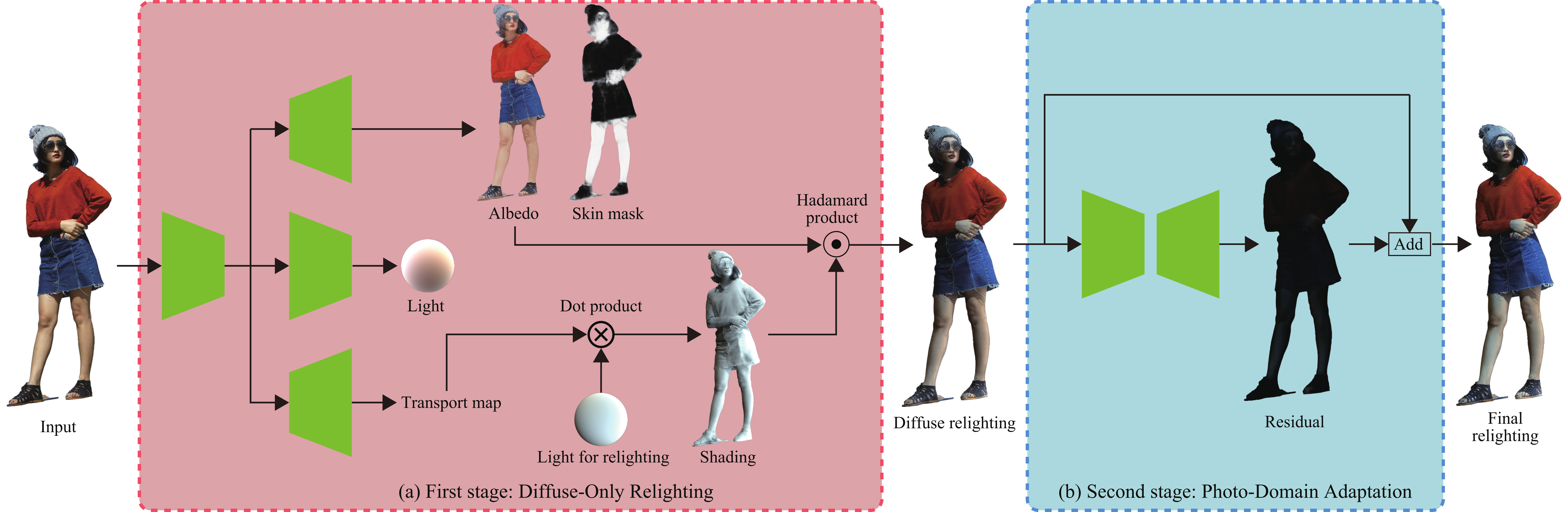}
\caption{\chkD{Overview of our two-stage relighting approach. The first-stage network infers diffuse-only relighting results, which is then refined in the second-stage network to handle non-diffuse reflection.}}
\label{fig:flow}
\end{figure*}

\section{Related Work}
\label{sec:RelatedWork}

\chkD{Because relighting has a long history of research, here we only introduce recent learning-based approaches that target single \chkE{human} images or videos as inputs. 
The previous studies differ in the targeted body parts (e.g., face only or full-body), training data types (e.g., 3D data or image only), and methodologies (e.g., inverse rendering or GAN inversion~\cite{xia2021gan}).} 

\paragraph*{Portrait relighting.}
\chkD{In the relighting literature, the term ``portrait'' means that the input images are face-only or headshot (i.e., hair and face with cloth up to chest) photos. Face-only methods~\cite{mofa,sfsnet,dsipr} often rely on the fitting of 3D morphable models~\cite{3dmm,10.1145/3395208} to grasp the face geometry.
Face relighting is also accomplished by transferring shading effects from reference images~\cite{mt,sthp}.
The recent approaches for headshots~\cite{sipr,drf,lpgfr,nlt,TotalRelighting} utilize the high-quality data captured using a special device called the \textit{Light Stage}~\cite{arfhf} (or similar ones~\cite{sipre,DBLP:journals/corr/abs-2104-00484}). The Light Stage can capture subjects lit by a single directional light from each light direction, yielding \textit{one-light-at-a-time} (OLAT) images. An OLAT image contains complicated lighting effects (including subsurface scattering and interreflection), and the superposition of OLAT images can represent lighting effects by environmental illuminations. The basic strategy of recent techniques is then to regress OLAT images directly or indirectly from the input headshots~\cite{sipr,drf,lpgfr,nlt,TotalRelighting} or videos~\cite{DBLP:journals/corr/abs-2104-00484}. The problem is that Light Stage data are not publicly available but accessible only to the self-sufficient research teams. As such, the Light Stage dataset used in each work is rather small (e.g., only 18~\cite{DBLP:journals/corr/abs-2104-00484} or 70 individuals~\cite{TotalRelighting}), resulting in poor generalization capability, particularly in handling cloth texture variations \chkF{(see, e.g., Section 8 in \cite{TotalRelighting})}. Furthermore, to the best of our knowledge, there are no \chkF{publicly-available} Light Stage datasets for full-body human figures, which are our target in this paper.}

\paragraph*{Full-body human relighting.}
\chkD{Relighting of full-body images or videos has various applications such as  human image cut-and-pasting~\cite{DBLP:journals/tog/XueADR12} and image-based virtual try-on~\cite{awesomeVTON} but has not been explored well so far. Kanamori and Endo~\cite{base} proposed a full-body human image relighting based on inverse rendering using convolutional neural networks (CNNs). The problem is that their method only supports diffuse relighting because their training dataset generated from 3D scanned human figures contains only diffuse textures. Our two-stage method adopts an improved version of their method in the diffuse-only first stage and handles non-diffuse relighting via domain adaptation to real photos in the second stage. We elaborate their method in Section~\ref{sec:Baseline}.}

\paragraph*{Domain adaptation.} 
\chkD{It is well known that supervised methods learning from synthetic data often suffer from the domain gaps between synthetic and real data. A remedy for this is domain adaptation.
Domain adaptation has been successfully integrated in various tasks such as 3D human pose estimation~\cite{DBLP:conf/mm/ZhangWKG19}, 3D object reconstruction~\cite{DBLP:conf/iccv/PinheiroRA19}, image dehazing~\cite{DBLP:conf/cvpr/ShaoLRGS20}, monocular depth estimation~\cite{DBLP:conf/cvpr/ZhaoFGT19}, semantic segmentation~\cite{DBLP:journals/ijcv/ZhengY21}, and object detection~\cite{DBLP:conf/eccv/SasagawaN20}.
To the best of our knowledge, our approach is the first attempt of domain adaptation for human image relighting.}

\paragraph*{GAN inversion.}
\chkD{An interesting, promising way of realistic face relighting is to utilize pre-trained GAN networks, namely, \textit{GAN inversion}~\cite{xia2021gan}. In GAN inversion, an input image is projected into the latent space of a pre-trained GAN network via optimization or an additional encoder, is manipulated in the latent space, and is then re-synthesized as a semantically manipulated yet realistic image. GAN inversion also allows us realistic face relighting~\cite{DBLP:journals/corr/abs-2103-07658,10.1145/3447648} without involving rendering process. Although pre-trained GAN networks for faces are publicly available~\cite{awesomeStyleGAN2}, those for full-body human images are not available and thus we cannot employ GAN inversion for our purpose.}

\chkF{A concurrent work by Lagunas et al.~\cite{sifhr} extends \chkG{the method by Kanamori and Endo}~\cite{base} by explicitly handling specular reflections with higher-order (i.e., 4th-order) spherical harmonics (SH) and a residual shading term. Their method is orthogonal to ours and could be further enhanced by our domain adaptation technique.}

\section{Brief Review of Our Baseline~\cite{base}}
\label{sec:Baseline}
\chkD{We adopt the network by Kanamori and Endo~\cite{base} as our baseline for the diffuse-only network used in the first stage. 
Their multi-task network is designed for the inverse rendering of full-body human images, and their three decoder branches extract an albedo map, light coefficients, and a light transport map respectively from each masked image.
The light coefficients and light transport maps are represented as coefficient vectors of second-order \chkF{SH}.
Unlike the previous methods~\cite{sfsnet,dsipr} that did not consider light occlusions, their method explicitly handles light occlusion at each pixel of the light transport map, yielding realistic shading, particularly in hollowed regions such as armpits, crotches, and garment wrinkles.
A drawback is that their method is limited to diffuse reflection because their training dataset is synthesized from 3D scanned human models, which usually contain only diffuse components as textures.} 

\chkD{We also point out two more problems in their method, namely, 1) the albedo-light color ambiguity and 2) the output blurriness.}
\begin{description}
\item[Albedo-light color ambiguity:] \chkD{The image decomposition in reverse rendering holds an intrinsic ambiguity; for example, from a masked photo of blue cloth, we cannot know whether a blue cloth is lit by a white light or a white cloth is lit by a blue light. 
As a result, although the multiplication of their albedo map and light is close to the input color, each color sometimes becomes strange, particularly if the ground-truth light is dissimilar to any of the training data.}
\item[Output blurriness:] \chkD{Although their method uses L1 total variation (TV) losses to enhance image edges, their albedo maps and shading maps (i.e., the multiplication of the light transport map and light) often lose details. 
For example,  the important texture details of faces, e.g., eyes, eyebrows, and lips in albedo maps, are smeared out.}
\end{description}
\chkD{We address these two problems in Section~\ref{sec:FirstStage}.}

\section{Method}
\label{sec:Method}



\chkD{Our method takes a single, masked human image as input and applies relighting under a new illumination.
We account for diffuse and non-diffuse reflection by a two-stage approach; we synthesize a diffuse-only image in the first stage (Section~\ref{sec:FirstStage}) and then compensate non-diffuse components in the second stage (Section~\ref{sec:SecondStage}).
Regarding the training datasets (Section~\ref{sec:Dataset}), the first-stage network uses synthetic images rendered from 3D scanned models. In contrast, the second-stage network relies on the outputs of the first stage and real human photos to account for non-diffuse residuals. 
Figure~\ref{fig:flow} shows the overview of our method.}

\chkD{Furthermore, we extend our method to human video relighting and demonstrate stable results with much less flickering using our illumination-conditioned deep video prior (Section~\ref{sec:VideoRelighting}).}

\begin{figure*}[tb]
\centering
\includegraphics[width=\linewidth]{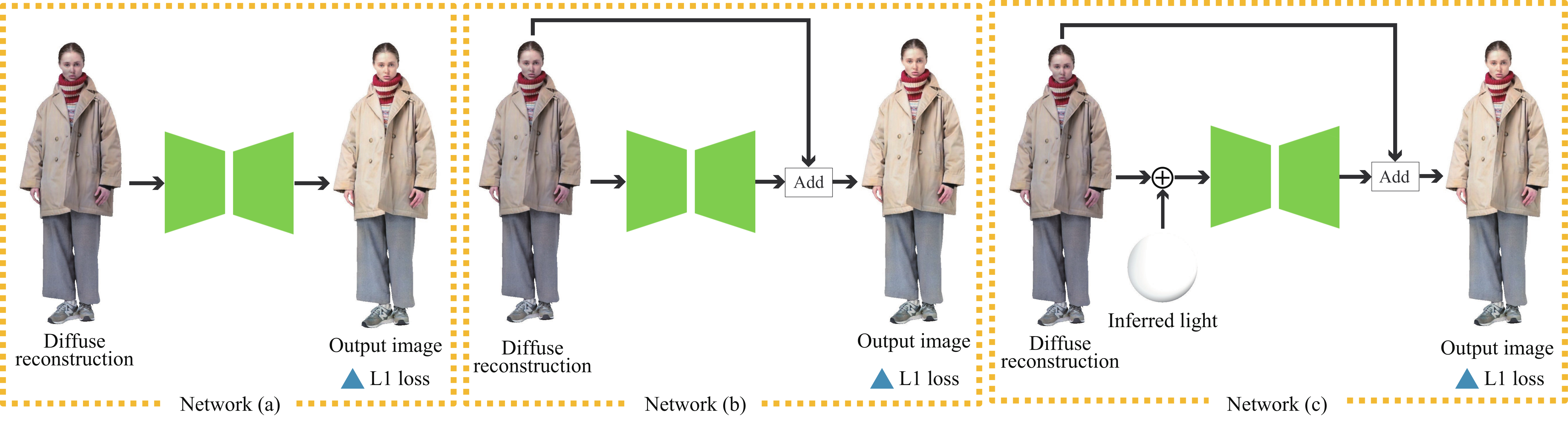}
\caption{\chkD{Three candidate network architectures for the second stage. We adopt (b) because of its best performance.}}
\label{fig:net123}
\end{figure*}

\subsection{First Stage: Diffuse-Only Network}
\label{sec:FirstStage}

\chkD{We adopt the \chkG{baseline} network architecture~\cite{base} with some modifications for the diffuse-only network in the first stage.
From a single human image, the diffuse-only network extracts an albedo map, light coefficients, and a light transport map, as explained in Section~\ref{sec:Baseline}.
We can reconstruct a diffuse-only version of the input image by multiplying the three components, i.e., dot products of the light coefficients and light transport map, followed by pixel-wise multiplication of the albedo map.  
The diffuse-only reconstruction is then fed to the second stage.}

\chkD{Here we propose solutions to address the two problems mentioned in Section~\ref{sec:Baseline} as follows:}
\begin{description}
\item[Solution to albedo-light color ambiguity:] \chkD{The strange output colors in the baseline method partially stem from their small light dataset; they only used 40 lights for training.
We thus increase the training data for lights from 40 to 2,760 by carefully selecting naturally-looking lights. 
Also, we disambiguate the colors of albedo maps and lights based on human-intrinsic color constancy\chkG{, inspired by Wang et al.~\cite{sipre}}; 
because skin colors are often stable in albedo maps, we integrate a mechanism to emphasize skin region \chkF{by inferring an albedo map and a skin mask simultaneously (see Section~\ref{sec:Training} for the implementation detail).}
Note that \chkG{Wang et al.} \chkF{extracts} face masks to emphasize face regions,
but face regions are often too small in our target images (i.e., full-body images).
In contrast, our skin mask covers a face and other regions such as arms and legs.}

\item[Solution to output blurriness:] \chkD{Instead of TV losses used in \chkG{the baseline}~\cite{base}, we introduce the spatially-frequency (SF) loss~\cite{sfloss} for optimizing albedo and shading maps to enhance their details. 
The SF loss is a weighted sum of L2 losses for images filtered by \chkE{the} normalized Laplacian-of-Gaussians (LoG) with different sigmas. \chkE{The normalied LoG kernel is defined as follows:
\begin{align}
    G_{\sigma}(\mathbf{x}) = - \frac{1}{2 \pi \sigma^2} \left( 2 - \frac{\norm{\mathbf{x}}^2}{\sigma^2}  \right) \exp \left( - \frac{\norm{\mathbf{x}}^2}{\sigma^2} \right), \label{eq:LoGKernel}
\end{align}
where $\sigma$ is the scale and $\mathbf{x}$ is the pixel position within the kernel.
For each subband $i$, we use different $\sigma_i \in \{ 0.6, 1.2, 2.4, 4.8, 9.6, 19.2 \}$ and weights $w_i \in \{ 600, 500, 400, 20, 10, 10 \}$ to put higher emphasis on higher-frequency subbands.
Let $A$ and $S$ be the ground-truth albedo and shading maps and $\tilde{A}$ and $\tilde{S}$ be their inferred counterparts.
The SF losses $\mathcal{L}^{albedo}_{SF}$ and $\mathcal{L}^{shading}_{SF}$ for albedo and shading maps are then defined as follows:
\begin{align}
    \mathcal{L}^{albedo}_{SF} = \frac{1}{C \, W \, H} \sum_i^6 w_i \, \norm{G_{\sigma_i} * A - G_{\sigma_i} * \tilde{A}}^2_2, \\
    \mathcal{L}^{shading}_{SF} = \frac{1}{C \, W \, H} \sum_i^6 w_i \,  \norm{G_{\sigma_i} * S - G_{\sigma_i} * \tilde{S}}^2_2,
\end{align}
where $*$ denotes convolution while $C$, $W$, and $H$ are the numbers of channels, width, and height of the maps, respectively.}
We observe that the SF loss improves high-frequency details \chkG{particularly} in the output albedo maps.
We demonstrate the resultant albedo maps in Section~\ref{sec:Results}.}
\end{description}

\subsection{Second Stage: Refinement Network}
\label{sec:SecondStage}

\renewcommand{\labelenumi}{(\alph{enumi})}

\chkD{In the second stage, the refinement network learns the differences between the masked photos of human figures and diffuse-only counterparts reconstructed in the first stage.
We adopt an encoder-decoder network with skip connections as the network architecture.
Specifically, we considered the following three architectures (see Figure~\ref{fig:net123}):}
\begin{enumerate}
\item \chkD{A network that directly generates an output containing both diffuse and non-diffuse reflections (Figure~\ref{fig:net123}, left),}
\item \chkD{A network that outputs the residuals from the diffuse-only reconstruction (Figure~\ref{fig:net123}, middle), and}
\item \chkD{A variant of (b) with inferred light information (Figure~\ref{fig:net123}, right).}
\end{enumerate}
\chkE{Among these candidates, (a) is the simplest network architecture to directly output non-diffuse residuals.
(b) is inspired by ResNet~\cite{resnet} so that the network outputs the difference between the ground-truth and the first-stage output.
(c) is to account for the light specified for relighting.}

\begin{table}[tb]
\caption{\chkE{Quantitative comparison of three network architecture for the second stage. The best value for each metric is in boldface.}}
\label{tab:2ndStageComparison}
\centering
\begin{tabular}{c|rrr} \hline
    & \multicolumn{1}{l}{RMSE $\downarrow$} & \multicolumn{1}{l}{SSIM $\uparrow$} & \multicolumn{1}{l}{LPIPS $\downarrow$}  \\ \hline \hline
(a) & 0.0419              & 0.990            & 0.0141          \\
(b) & \textbf{0.0417}     & \textbf{0.996}   & \textbf{0.0059} \\
(c) & 0.0428              & 0.990            & 0.0136          \\  \hline
\end{tabular}
\end{table}

\chkE{Table~\ref{tab:2ndStageComparison} shows the result of quantitative comparison among the three architectures.}
\chkG{We can see network (b) was the best, and thus we adopt (b) for the refinement network.
One might consider that network (c) should have been better due to the light integration.
However, in reality (c) was inferior, probably due to the erroneous lights inferred by our baseline network; note that there are no ground-truth lights for real photos.}
\renewcommand{\labelenumi}{(\arabic{enumi})}

\subsection{Network Training}
\label{sec:Training}

\chkD{Here we summarize how to train our networks. 
We first train the diffuse-only network in the first stage using a synthetic dataset. 
After freezing the network parameters, we apply the pre-trained diffuse-only network to real photos of human figures to calculate diffuse-only reconstructions.
We then train the refinement network to learn the residuals between the real photos and diffuse-only counterparts.}

\chkF{We explain our network architectures.  For the first stage, we borrowed the encoder-decoder network by Kanamori and Endo~\cite{base} (see Section 5 in their paper); it has three decoders for albedo, light transport, and light\chkG{, respectively}. Their encoder has six convolutional layers, a residual block, and skip-connections with the decoder branches for albedo and light transport maps. We modified their network as follows. We replaced each transposed convolution in the decoders with upsampling and convolutional layers to avoid checkerboard artifacts. In the albedo decoder, we modified the final layer so that it generates not a three- (i.e., RGB albedo map) but a four-channel output (i.e., albedo map and skin mask). For the second stage, we used a U-net-like architecture whose encoder and single-branch decoder are almost identical to those used in the first stage, except for the numbers of input/output channels.}

\chkD{We used the following loss functions for training.
\chkF{The loss function $\mathcal{L}^{1st}$ for the first stage is defined as:
\begin{align}
 \mathcal{L}^{1st} \! = \!\! \chkG{\sum_{r \in \mathcal{R}} \!\! \mathcal{L}^{reconst.}_{r}} \! + \!\! \chkG{\sum_{s \in \mathcal{S}} \!\! \mathcal{L}^{shading}_{s}} \!
 + \mathcal{L}^{albedo}_{SF} \!\! + \mathcal{L}^{shading}_{SF} \!\! + \mathcal{L}^{focal} \!\! ,
\end{align}
where \chkG{$\mathcal{L}^{reconst.}_{r}$} and \chkG{$\mathcal{L}^{shading}_{s}$} are the reconstruction and shading losses used in \chkG{the baseline}~\cite{base}\chkG{; $\mathcal{R}$ and $\mathcal{S}$ are sets of subscripts for ten types of reconstruction losses (i.e., three unary losses for albedo map, light transport map, and light, and seven patterns of triple products of albedo map, light transport map, and light) and three types of shading losses (i.e., three patterns of double products of light transport map and light), respectively.
Please refer to the paper~\cite{base} for more details.}} 
\chkF{As} explained in Section~\ref{sec:FirstStage}, we replace \chkF{their} TV losses with \chkF{our} SF losses \chkF{$\mathcal{L}^{albedo}_{SF}$ and $\mathcal{L}^{shading}_{SF}$}.
We also added a focal loss \chkF{$\mathcal{L}^{focal}$~\cite{sipre}} for the skin mask.
\chkF{Each loss in $\mathcal{L}^{1st}$ has the same weight (i.e., 1).}
In the second stage, we \chkF{only} used an L1 loss between the final outputs and real photos.}


\subsection{Flicker-Tolerant Human Video Relighting}
\label{sec:VideoRelighting}

\chkD{We also propose a simple extension of our method to human video relighting.
We apply our method for each frame of an input video separately.
However, as is often the case with frame-by-frame image processing, this straightforward approach suffers from flickering because we cannot guarantee temporal consistency.
To remove the flickering artifacts, we adopt the \textit{deep video prior} (DVP) proposed by Lei et al.~\cite{dvp}.
Inspired by the finding~\cite{dip} that CNNs learn low-dimensional manifolds faster than noise, Lei et al. extended this idea to the temporal dimension to reduce temporal noise (i.e., flickering), resulting in a simple formulation with excellent noise reduction. 
\chkE{Unlike previous approaches~\cite{DBLP:journals/tog/BonneelTSSPP15,DBLP:conf/eccv/LaiHWSYY18}, DVP does not need additional training datasets or estimating optical flow. To apply} DVP, we train an untrained network \chkF{(of the same architecture as the second stage)}
to \chkF{fit} the frames of an input video \chkF{(i.e., the original video before applying relighting) to those of the relit video containing flickering,}
\chkF{only} with the perceptual loss.
\chkF{We then} terminate the training \chkF{manually} before noise appears in the result.}

\chkD{We confirmed that DVP removes flickering excellently with static illuminations but causes a problem with dynamic illuminations.
If we animate an illumination (by, e.g., rotating it), flickering artifacts are removed, but the illumination looks as if it were static; we cannot see time-varying shading effects due to the dynamic illumination. 
\chkI{This is probably because some input frames are similar, but their ground-truth outputs are different from each other; in such a case, the network has to infer different outputs from similar inputs by learning one-to-many correspondence, which is difficult to learn.}
To account for dynamic illumination, we condition the network by feeding light information in addition to each frame.
Specifically, we vectorize the SH coefficients of the illumination for each frame and replicate it for each pixel in the frame.
We then concatenate the input frame and the tiled SH coefficient vectors to feed the network.
}
\chkI{The SH coefficients serve as a powerful cue to distinguish similar input frames, enabling time-varying shading effects due to dynamic illuminations.}


\section{Dataset}
\label{sec:Dataset}

\chkD{We \chkE{gathered 541} 3D scanned models in standing poses from several commercial websites for the first stage and used \chkE{521} for training and \chkE{20} for test.
For each 3D model, we rendered a binary mask, albedo map, and light transport map at the resolution of $1024 \times 1024$, as done in \chkG{the baseline}~\cite{base}.
We collected HDR environment maps from HDRI Heaven\footnote{\webLink{https://hdrihaven.com/}} and \chkE{calculated} their SH coefficients \chkE{after rotating them at $36^\circ$ degree intervals. We used the resultant 2,760 lights for training. For test, we used the 10 lights provided by the authors~\cite{base}.
To prepare ``ground-truth'' skin masks for our synthetic human image dataset, we applied a semantic segmentation network trained with the \textit{ModaNet}~\cite{DBLP:conf/mm/ZhengYKP18} dataset, which does not contain ``skin'' labels.
We extracted skin masks as the intersections of our binary masks and ``background'' regions in the ModaNet dataset.}}

\chkD{For the second stage, we collected 9,152 photos of standing fashion models from a website of an apparel company and used 8,900 photos for training and 252 for test.
We extracted binary masks for the real photos using a commercial service\chkF{\footnote{\webLink{https://remove.bg/}}}. 
The photos and binary masks were uniformly resized to $1024 \times 1024$ with appropriate paddings.
}


\begin{table*}[tb]
\centering
\caption{\chkD{Quantitative ablation study for our first stage (plus \cite{sfsnet}) with synthetic data. The best scores are in boldface.}}
\label{tab:CGReconstruction}
\begin{tabular}{l|rrrr|rrr|rrr} \hline
                 & \multicolumn{4}{c|}{RMSE $\downarrow$}   & \multicolumn{3}{c|}{SSIM $\uparrow$}   & \multicolumn{3}{c}{LPIPS $\downarrow$}   \\
\textbf{}        & \multicolumn{1}{l}{Albedo} & \multicolumn{1}{l}{Light} & \multicolumn{1}{l}{Shading} & \multicolumn{1}{l|}{Reconst.} & \multicolumn{1}{l}{Albedo} & \multicolumn{1}{l}{Shading} & \multicolumn{1}{l|}{Reconst.} & \multicolumn{1}{l}{Albedo} & \multicolumn{1}{l}{Shading} & \multicolumn{1}{l}{Reconst.} \\  \hline  \hline
{[}SKCJ18{]}     & 0.101                      & 0.290                     & 0.274                       & 0.0581                             & 0.982                      & 0.813                       & 0.994                              & 0.0280                     & 0.0888                      & 0.0102                             \\
{[}KE18{]}       & 0.104                      & 0.265                     & 0.207                       & 0.0600                             & 0.979                      & 0.969                       & 0.992                              & 0.0447                     & 0.0444                      & 0.0177                             \\
Ours w/o SF loss & 0.0996                     & 0.142                     & 0.120                       & 0.0428                             & 0.977                      & 0.979                       & 0.992                              & 0.0494                     & 0.0413                      & 0.0228                             \\
Ours             & \textbf{0.0932}            & \textbf{0.131}            & \textbf{0.112}              & \textbf{0.0377}                    & \textbf{0.986}             & \textbf{0.982}              & \textbf{0.996}                     & \textbf{0.0229}            & \textbf{0.0292}             & \textbf{0.00650}  \\ \hline                 
\end{tabular}
\end{table*}

\begin{table}[tb]
\centering
\caption{\chkD{Quantitative evaluation of output images reconstructed from real photos. The best scores are in boldface.}}
\label{tab:PhotoReconstruction}
\begin{tabular}{l|lll} \hline
     & RMSE $\downarrow$    & SSIM $\uparrow$           & LPIPS $\downarrow$    \\ \hline \hline
{[}SKCJ18{]} & 0.0749          & 0.988          & 0.0130          \\
{[}ZHSJ19{]} & 0.0813          & 0.983          & 0.0201          \\
{[}KE18{]}   & 0.0783          & 0.989          & 0.0197          \\
Ours         & \textbf{0.0417} & \textbf{0.996} & \textbf{0.0059} \\ \hline
\end{tabular}
\end{table}

\begin{figure*}[tb]
\centering
\includegraphics[width=1.\linewidth]{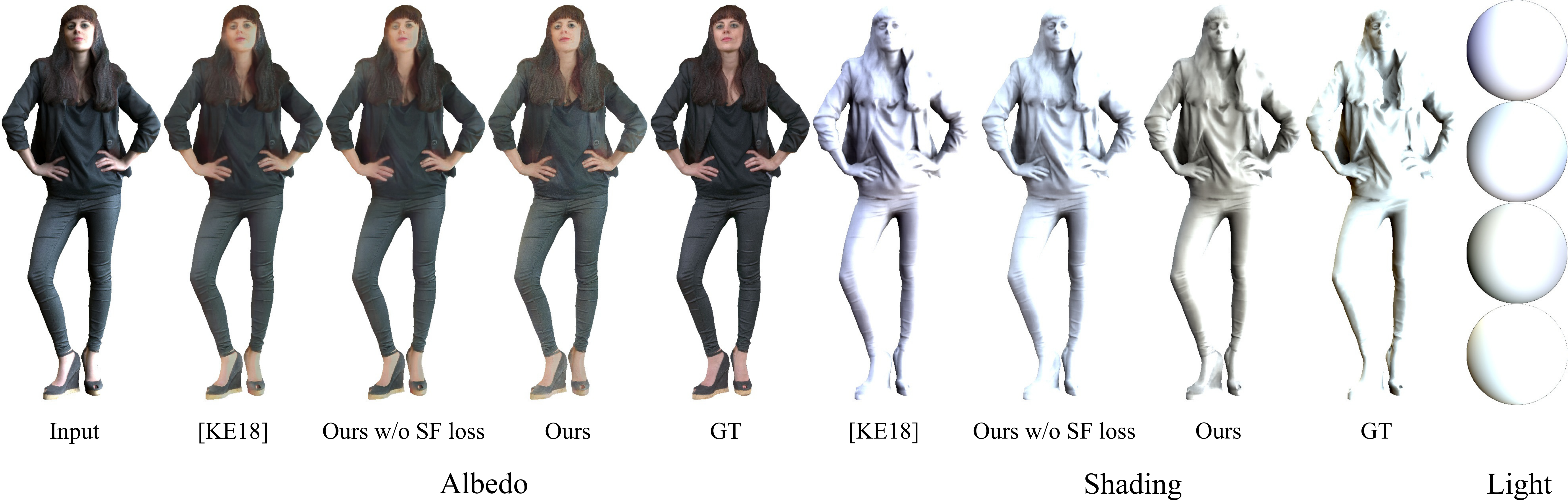}
\caption{\chkE{Qualitative evaluation for our first stage with synthetic data. Lights are sorted in the same order as albedo and shading maps, from top to bottom.}}
\label{fig:abl_3d}
\end{figure*}

\begin{figure*}[tb]
\centering
\includegraphics[width=1.\linewidth]{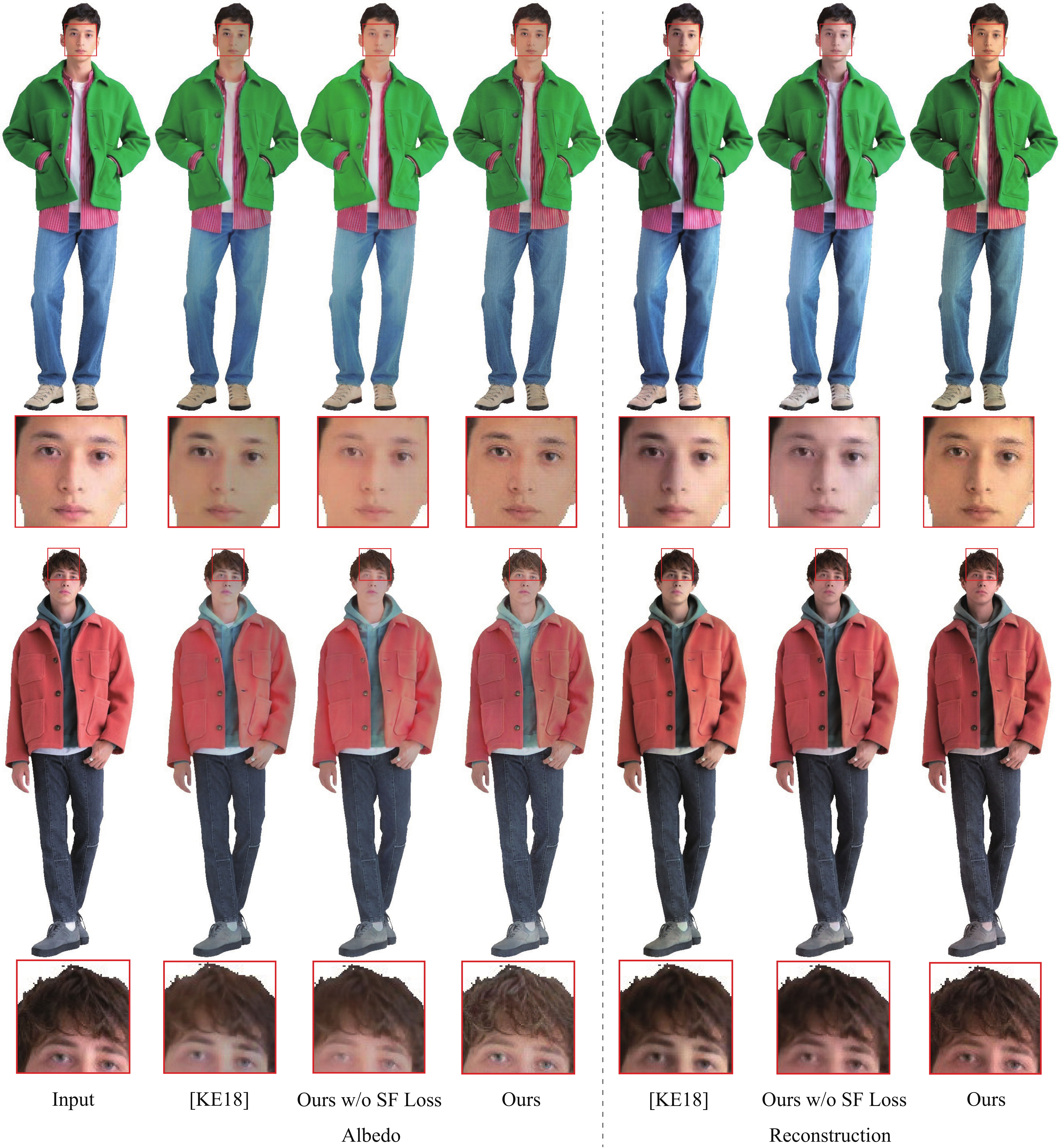}
\caption{\chkE{Qualitative evaluation for our first stage with real photos.}}
\label{fig:abl_real}
\end{figure*}

\begin{figure*}[tb]
\centering
\includegraphics[width=\linewidth]{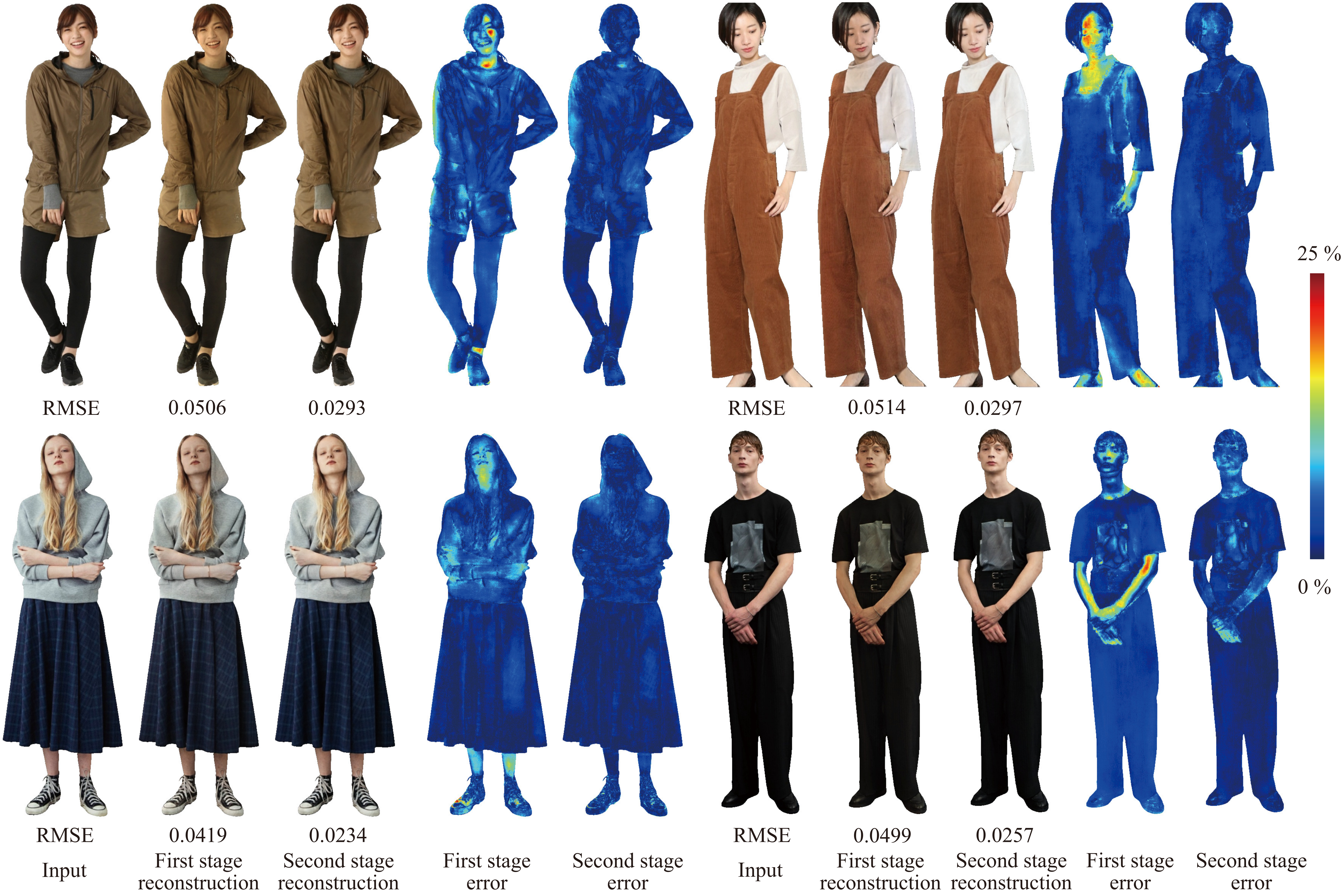}
\caption{\chkD{Comparison of our first- and second-stage outputs. The blueish images encode per-pixel absolute errors.}}
\label{fig:ql_recon}
\end{figure*}

\begin{figure*}[tb]
\centering
\includegraphics[width=1.\linewidth]{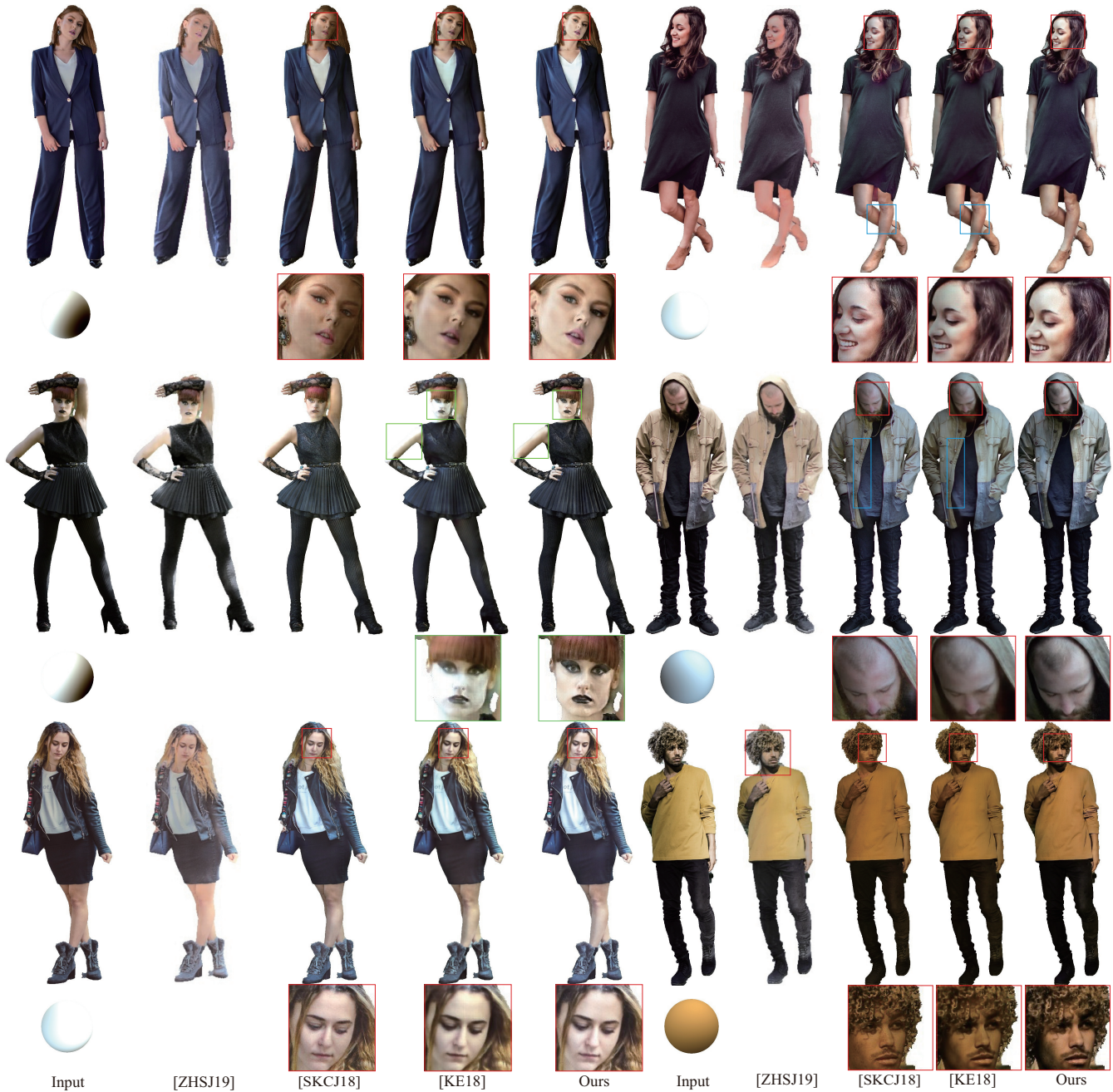}
\caption{\chkD{Comparison of relighting results. The rectangles emphasize skin highlights (red), occluded regions (blue), and albedo differences (green).}}
\label{fig:ql_relit}
\end{figure*}


\begin{figure}[tb]
\centering
\includegraphics[width=\linewidth]{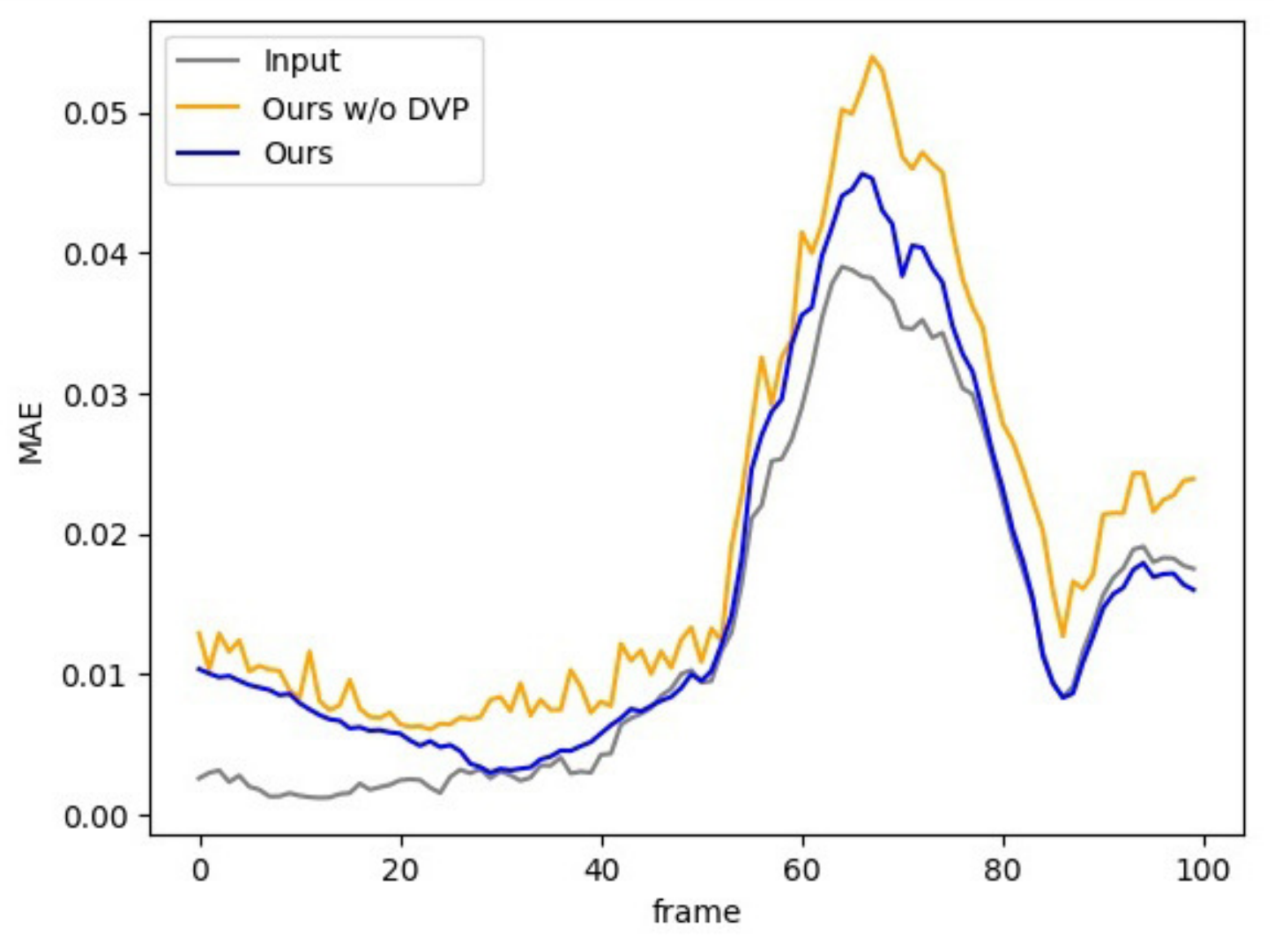}
\caption{\chkD{Graphs of temporal differences between two consecutive video frames.}}
\label{fig:FlickerGraph}
\end{figure}

\section{Results}
\label{sec:Results}

\chkD{We implemented our method using Python and PyTorch and conducted training and inference on NVIDIA GeForce GTX 1080 Ti. 
We used Adam as an optimizer \chkE{for the first stage and RAdam for the second stage}, setting the exponential decay rates for the moment estimates as \{0.5, 0.999\}.
We scheduled the learning rates in the range of \chkE{[0.001, 0.0001]} using the cosine annealing within \chkE{20} epochs per cycle.
Our batch size was one.
The computational time for one-epoch training was about 30 minutes when we used one GPU to process $1024 \times 1024$ images.
We terminated the training at 180 epochs, where the learning curves reached a plateau. 
The time for testing a $1024 \times 1024$ input image was about 0.18 seconds.}

\chkD{To evaluate our method, we conducted the following experiments:
\begin{enumerate}
\item Ablation study of the first stage,
\item Difference evaluation between the first and second stages,
\item Comparison with existing relighting methods, and
\item Evaluation of human video relighting.
\end{enumerate}
As the evaluation metrics, we calculated RMSE within binary masks, SSIM and LPIPS~\cite{DBLP:conf/cvpr/ZhangIESW18} within bounding boxes of masks to exclude pixels out of masks or bounding boxes.
Because real photos do not have ground-truth relighting results, we measured the errors between input images and reconstructed images for quantitative evaluations.
Our results in this paper were generated from the test data only.}

\subsection{Ablation Study of the 1st Stage}
\label{sec:1stAblation}

\chkD{We evaluated the effectiveness of our improvements (i.e., integration of the skin mask and SF loss, explained in Section~\ref{sec:FirstStage}) over the original diffuse-only baseline~\cite{base}.
Specifically, we compared 1) the baseline~\cite{base}, 2) ``Ours w/o SF loss'' (i.e., skin mask only), and 3) ``Ours'' (i.e., with both skin mask and SF loss).
\chkF{We trained the baseline~\cite{base} with the same training data (i.e., both lights and 3D models) as ours for a fair comparison.}
Tables~\ref{tab:CGReconstruction} and \ref{tab:PhotoReconstruction} show the results of quantitative evaluations with synthetic data and real photos as inputs, respectively.
Both tables demonstrate that ``Ours'' recoded the best scores in all metrics.
Figures~\ref{fig:abl_3d} and \ref{fig:abl_real} show the qualitative comparisons with synthetic data and real photos as inputs, respectively.
We can see that skin masks suppress highlights in skin regions and thus help recovering natural skin colors in the albedo maps.
With the SF loss, high-frequency details are extracted in the albedo maps as well as the reconstructed images.}

\subsection{Difference Evaluation between the 1st and 2nd Stages}
\label{sec:1st2ndDiff}

\chkD{Figure~\ref{fig:ql_recon} shows the reconstructed results of the first (i.e., diffuse-only) and second (i.e., refinement) stages as well as their differences.
The RMSE scores under the reconstructed images are almost halved from the first to the second stages.
The blueish images visualize per-pixel absolute differences.
The first-stage results have large errors particularly in skin regions whereas the errors in the second stage are reduced substantially, which implies that \chkF{the glossiness (e.g., skin highlights) is} recovered in the second stage.}

\subsection{Comparison with Existing Relighting Methods}
\label{sec:ComparisonExistingMethods}

\chkD{We compared our method with three existing relighting methods~\cite{sfsnet,dsipr,base}.
Both of \cite{sfsnet,dsipr} were originally proposed for face relighting.
We trained \cite{sfsnet} with our synthetic dataset.
In the training of \cite{dsipr}, we had to fabricate normal maps for photographed people, which we synthesized using a recent method for single-image 3D human reconstruction~\cite{pifuhd}.
We then trained our own implementation of \cite{dsipr} with real photos obtained from an apparel website and lights inferred using \chkG{the baseline}~\cite{base}.}

\chkD{Table~\ref{tab:PhotoReconstruction} demonstrates that our method outperforms the others in all metrics w.r.t. the reconstruction errors.
Figure~\ref{fig:ql_relit} shows the qualitative comparison, where the rectangles emphasize skin highlights (red), occluded regions (blue), and albedo differences (green).
Overall \chkG{the method by} \chkF{Zhou et al.}~\cite{dsipr} did not work well probably because \chkG{it} could not learn light variations because fashion models in apparel websites are almost always front-lit.
As pointed out \chkG{by Kanamori and Endo}~\cite{base}, \cite{sfsnet} makes occluded regions (blue rectangles) unnaturally bright because it ignores light occlusion.
\cite{base} generates blurry albedo maps (red rectangles) with possibly wrong colors (green rectangles).
Ours yields more natural skin colors with highlights (red rectangles).}

\chkA{To further validate our method, we \chkF{conducted} a user study \chkG{based on the Friedman test~\cite{siegel:b88}}.
Using six sets of relighting results shown in our paper, we requested 20 subjects to rank four \chkG{methods (i.e., \cite{dsipr}, \cite{sfsnet}, \cite{base}, and ours)} for each set \chkG{via an online survey using the Google Forms}, \chkG{asking} the naturalness of images relit under given lights. 
\chkG{The display order of the four methods was shuffled for each set.}
Consequently, we obtained 120 (= \chkF{$6 \times 20$}) samples.
\chkG{We summed up each sample's rank (i.e., 1, 2, 3, or 4) for each method.}
\chkF{The rank sums \chkG{(the smaller the better)} of \chkG{the four methods}
are 417, 314, 297, and 172, respectively, which means ours is the best.
We confirmed that the ranking has a statistically significant difference ($p < 0.001$).}
Furthermore, post-hoc tests revealed that ours outperforms the others statistically significantly.}

\subsection{Evaluation of Human Video Relighting}
\label{sec:VideoRelightingResults}

\begin{figure*}[tb]
\centering
\includegraphics[width=0.8\linewidth]{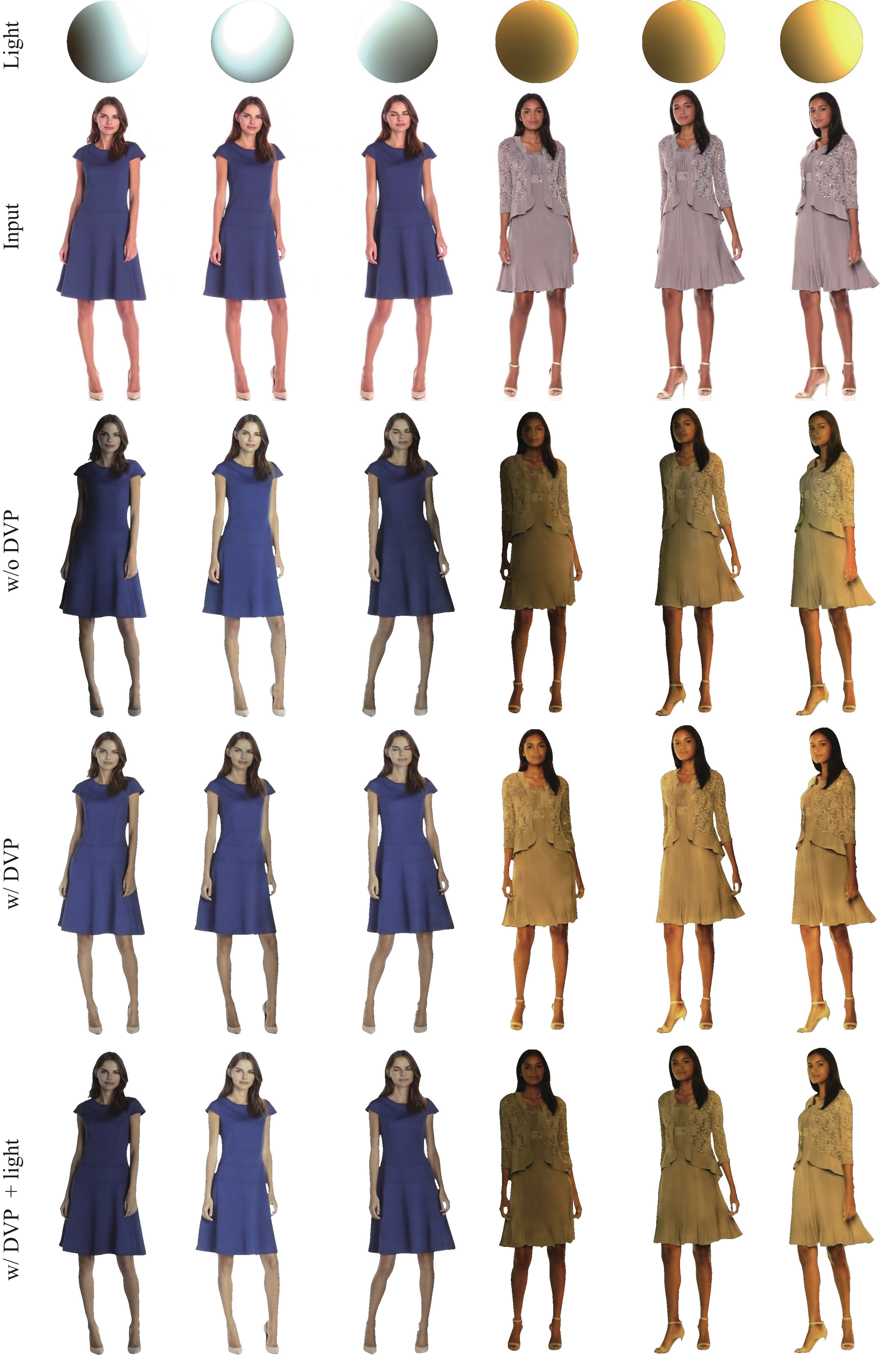}
\caption{\chkD{Relighting results for human videos. Please watch the accompanying video for the details.}}
\label{fig:ql_relit_video}
\end{figure*}

\chkD{We obtained \chkG{the} input videos \chkG{of Zablotskaia et al.~}\cite{fvd}, 
and then applied our method in the following three different ways:
\begin{description}
\item[``w/o DVP'':] Na\"{i}vely applying our method frame by frame,
\item[``w/ DVP'':] Simply applying DVP to our initial results, and
\item[``w/ DVP + light'':] Applying DVP with light conditioning.
\end{description}
Figure~\ref{fig:ql_relit_video} shows the resultant frames with two video sequences.
\textbf{Please watch the accompanying video to see the video outputs.}
When we applied our method without DVP, there were heavy flickering artifacts in the outputs.
When we employed DVP, the flickering artifacts were greatly reduced.
However, DVP stabilizes the output too strongly that it even cancels the time-varying shading effects by dynamic illuminations.
Consequently, the females subjects look as if they were lit by static lights.
With our light-aware DVP, we can account for dynamic illumination, while reducing flickering greatly.
The graphs in Figure~\ref{fig:FlickerGraph} plot the differences between two consecutive frames of the output video sequences (Figure~\ref{fig:ql_relit_video}, left) with a dynamic illumination.
Specifically, the differences are calculated as mean absolute error (MAE) within intersections of masks in two consecutive frames.
Although this na\"{i}ve MAE measurement is not the best way because we do not consider correspondences among frames, we can see the tendency that our light-aware DVP (the orange graph) yields lower errors than without DVP (the blue graph).}

\section{Limitations}
\label{sec:Limitations}

\begin{figure}[tb]
\centering
\includegraphics[width=\linewidth]{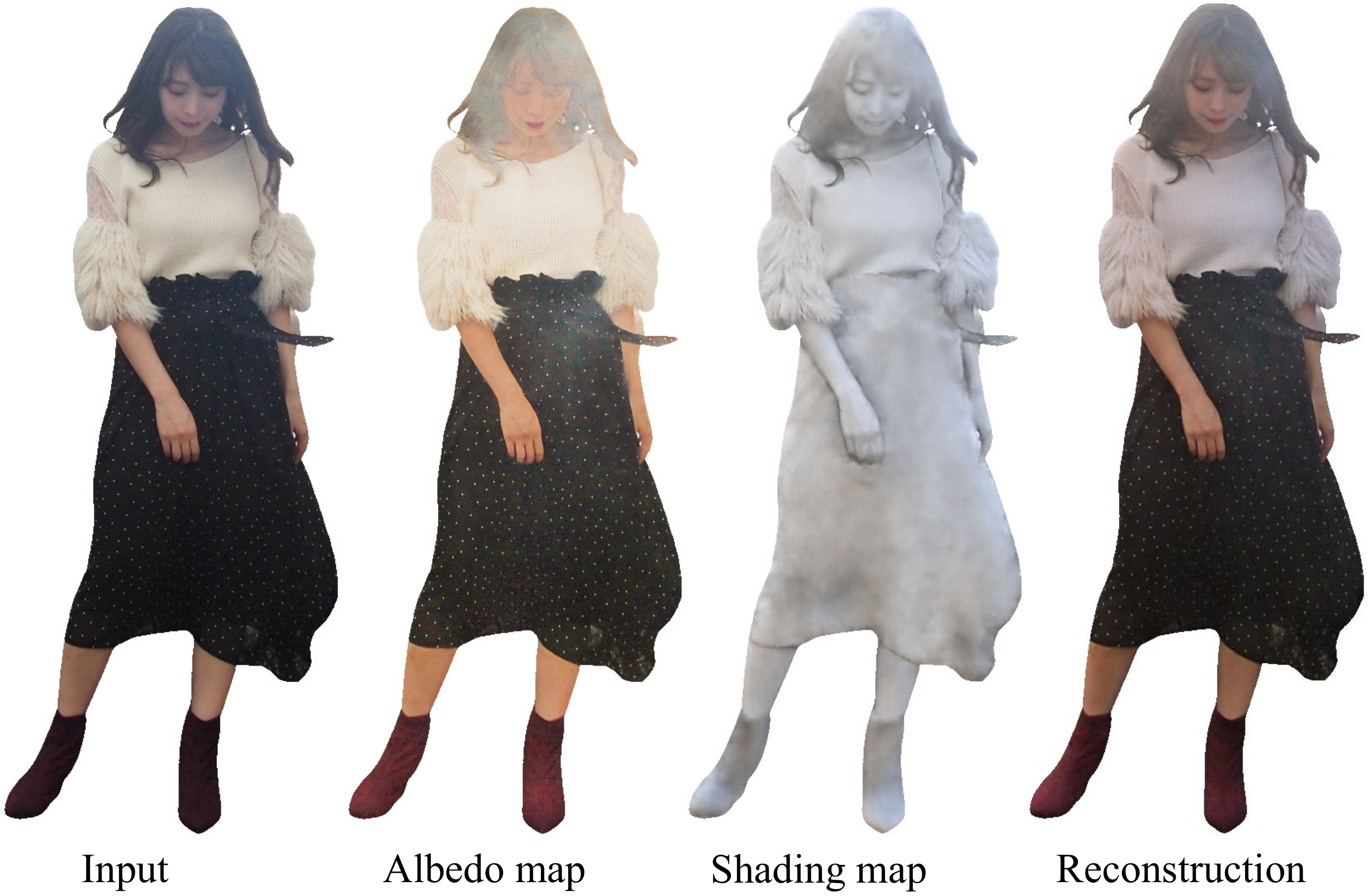}
\caption{Failure case of albedo-shading decomposition in the first stage.}
\label{fig:Failure}
\end{figure}

\chkD{Although our method can handle non-diffuse relighting, it inherits the limitations of the diffuse-only baseline~\cite{base}. 
If texture patterns are complicated or lighting conditions are extraordinary, the albedo-light color ambiguity is hard to resolve.
Consequently, albedo and shading maps might have wrong colors (see Figure~\ref{fig:Failure} as an example), causing poor relighting.
To address this issue, we would like to provide richer information to the networks; the light inference will be improved if we feed not only foreground images but also background images.
For better albedo inference, a larger training dataset will help networks learn richer albedo variations.}

\chkD{Our light-conditioned DVP greatly reduces flickering under dynamic illuminations but does not completely eliminate it.
We hypothesize that this is because the light information provided by SH coefficients is insufficient.
We would like to explore a better approach in future.}

\section{Conclusions}

\chkD{In this paper, we have proposed a two-stage supervised method for relighting full-body human images or videos, while enhancing generalization capability and reducing the synthetic-to-real domain gap.
The first stage infers diffuse-only relighting better than the baseline~\cite{base}, thanks to skin masks~\cite{sipre} and the SF loss~\cite{sfloss}.
The second stage compensates the \chkG{synthetic}-to-real gap by inferring the differences between diffuse-only relighting and real photos.
We further accomplished stable human video relighting by conditioning the deep video prior~\cite{dvp} with light information.
As a result, we can achieve high temporal coherence even under dynamic illumination. 
}

\chkF{There are several directions for future work.
Although we did not compare our method with Light-Stage-based methods~\cite{sipr,sipre,TotalRelighting} due to the lack of publicly-available full-body image datasets, we would like to confirm their advantage (i.e., the ability to account for arbitrary illumination effects such as hard shadows, glossy reflections, subsurface scattering, and inter-reflection) in our setting. 
We also would like to account for the background image for better light inference.}


\section*{Acknowledgements}

We thank the anonymous reviewers for their fruitful comments.
The real photograph dataset is by courtesy of ZOZO, Inc.
This work was supported by JSPS KAKENHI (Grant-in-Aid for Scientific Research (B)) Grant Number JP19H04130.


\printbibliography                

\end{document}